\documentclass[10pt,aps,twocolumn,nofootinbib]{revtex4-1}
\usepackage{amsfonts}
\usepackage{mathrsfs}
\usepackage{amsmath}
\usepackage{amssymb}
\usepackage[normalem]{ulem}
\usepackage{extarrows}
\usepackage{slashed}
\usepackage{isodateo}
\usepackage{graphicx} 
\usepackage{xcolor}
\usepackage[bookmarksnumbered=true,bookmarksopen=true]{hyperref}
 \hypersetup{colorlinks,%
             linkcolor=[rgb]{0,0.3,0.6}, %
             citecolor=[rgb]{0,0.3,0.6}, %
             urlcolor=[rgb]{0,0.3,0.6}}%
\renewcommand{\rm}{\mathrm} 
\begin{document}

\title{Dynamical Dark Energy in light of DESI BAO and Full-Shape Data}
\author{Quan Zhou}
\email{quan.zhou@stu.cqu.edu.cn}
\affiliation{Department of Physics, Chongqing University, Chongqing 401331, China}
\author{Sibo Zheng}
\email{sibozheng.zju@gmail.com}
\affiliation{Department of Physics, Chongqing University, Chongqing 401331, China}

\begin{abstract}
Recently, the DESI BAO data has reported a preference of dynamical dark energy (DDE) over the $\Lambda$CDM.
Apart from the BAO data, the DDE model should be also sensitive to low-redshift measurements of
matter power spectrum.
In this study, we address this point by combining the DESI Y1 data about the matter power spectrum, extracted from the DESI Full-Shape data, with the DESI DR2 BAO data among others.
After building the DESI Y1 likelihood,
we carry out a Markov Chain Monte Carlo analysis, showing that the constraints on $w_0$ and $w_{a}$ with DESI Y1 data included are improved over those without it for three different datasets widely considered,
especially in the case of DESY5 sample.
\\
\\
\end{abstract}
\date{December 8, 2025}
\maketitle

\section{Introduction}
Dark Energy Spectroscopic Instrument (DESI) is the first Stage-IV galaxy survey in operation to probe dark energy, neutrino mass and the expansion history of Universe.
Combing cosmic microwave background (CMB) and supernovae (SNe) data,
DESI measurements of baryon acoustic oscillation (BAO)
have pointed to a preference of dynamical dark energy (DDE) model over a constant dark energy.
This preference increases from $2.5-3.9\sigma$ level in the DESI DR1 data \cite{DESI:2024mwx} to more than $\sim 4\sigma$ in the DESI DR2 data \cite{DESI:2025zgx}. 

The DESI data has initiated extensive studies on the DDE model.
Refs.\cite{Zhong:2025gyn,DESI:2024kob,Gialamas:2024lyw,Malekjani:2024bgi,Roy:2024kni} considered the consistency of the DESI data with the other existing datasets.
Refs.\cite{Yang:2024kdo,Tada:2024znt,Li:2024qus,Liu:2025myr} carried out the implications of DESI data to theoretical models.
Lastly,  a few works such as Refs.\cite{Huang:2025som,Ong:2025utx} argued potential bias and systematic errors, 
pointing out that the current evidence for the DDE model is not yet conclusive.

In this work, we attempt to place tighter constraints on the DDE model by combing the DESI R2 BAO data and low-redshift measurements of matter power spectrum \cite{Tegmark:2002cy,Tegmark:2008au,Chabanier:2019eai} being sensitive to the DDE, 
which have been updated by the DESI Full-shape data \cite{DESI:2024hhd, DESI:2025xzu} recently. 
Refs.\cite{DESI:2024hhd, DESI:2025xzu} used Full-Modeling techniques to constrain the DDE model.
This method computes a linear matter power spectrum employing a Boltzmann code.
The derived spectrum serves as the basis of generating the non-linear redshift-space galaxy power spectrum, 
which is then used to fit observational data. Several recent studies have shown that directly fitting the full power spectrum signal or correlation function can place tighter constraints on relevant cosmological parameters compared to compression methods \cite{Ivanov:2019pdj,DAmico:2019fhj,Chen:2022jzq,Simon:2022csv,Chudaykin:2022nru} such as in the BAO analysis.

Concretely speaking, we do not consider SDSS DR7 LRG data \cite{Reid:2009xm} according to the argument of \cite{DESI:2025xzu} and eBOSS DR 14 Ly$\alpha$ forest data \cite{eBOSS:2017pfi} which has a $\sim 5\sigma$ tension \cite{Rogers:2023upm} with the Planck data about the matter power spectrum.

The rest of this paper is organized as follows. 
In Sec.\ref{datasets} we introduce the existing datasets considered in this paper,
where we present details to construct the DESI Y1 likelihood in terms of \cite{DESI:2025xzu} 
and mention model-dependent uncertainty related to this dataset. 
Sec.\ref{results} presents the Markov Chain Monte Carlo (MCMC) analysis of the DDE model under the Chevallier-Polarski-Linder (CPL) \cite{Chevallier:2000qy,Linder:2002et} parametrization, 
where we highlight improved constraints on the DDE model parameters over the previous results in \cite{DESI:2025zgx}, 
using the publicly available code CLASS \cite{Lesgourgues:2011re,Blas:2011rf} and COBAYA \cite{Torrado:2020dgo,cobaya2019}.
Finally, we conclude in Sec.\ref{con}.

\section{Datasets}
\label{datasets}
We consider the following datasets to place constraints on the DDE model. 
\begin{itemize}
\item Planck: the Planck PR4 CamSpec NPIPE likelihood \cite{Rosenberg:2022sdy} for the $\rm{high}\text{-}\ell$ TTTEEE spectrum, the $\rm{low}\text{-}\ell$ TT (commander) \cite{Planck:2019nip} and EE (sroll2) \cite{Pagano:2019tci} likelihoods, 
and combination of Planck and ACT DR6 lensing likelihoods \cite{ACT:2023kun,ACT:2023dou,Carron:2022eyg}.
\item BAO:  the DESI DR2 BAO \cite{DESI:2025zgx} measurements which encompass the BAO results for all tracers.
\item SNe: the PantheonPlus \cite{Brout:2022vxf}, Union3 \cite{Rubin:2023jdq}, and the Dark Energy Survey Year 5 (DESY5) \cite{DES:2024jxu} samples.
\item DESI Y1: we use the DESI Y1 data \cite{DESI:2025xzu},
which reconstructed the matter power spectrum $P_{m}(k)$ at redshift $z=0$ in terms of the DESI DR1 Full-Shape data \cite{DESI:2024aax, DESI:2024jxi}, 
to create the DESI Y1 likelihood class in Cobaya.
\end{itemize}

In the DESI Y1 likelihood analysis,
 we use the following definition to estimate the value of 
\begin{eqnarray}\label{chi}
\chi^{2}=\sum_{i} \left(\frac{P^{\rm{DDE}}_{m}(k_{i})-P^{\rm{exp}}_{m}(k_{i})}{\sigma_{i}^{2}}\right)^{2},
\end{eqnarray}
where $\sigma_{i}$ represents the independent error between each data point as seen in Fig.6 of \cite{DESI:2025xzu}.
We run the chains using the Metropolis-Hastings algorithm and consider them being converged when the Gelman-Rubin criterion  $R-1 < 0.01$ \cite{Gelman:1992zz} is satisfied.
We obtain statistics for the chains and plots with Getdist \cite{Lewis:2019xzd}.

Regarding uncertainties in the DESI Y1 data, 
we remind the reader that the DDE model parameters enter into the computations of the data points, 
as firstly mentioned in \cite{Chabanier:2019eai}. 
We determine the baseline values of $P^{\rm{DDE}}_{m}(k)$ related to $P^{\rm{exp}}_{m}(k_{i})$ \cite{DESI:2025xzu} in terms of the best-fit values of cosmological parameters extracted from the MCMC analysis of the DDE model with respect to an explicit dataset of Planck$+$DESI DR2 BAO$+\{\rm{PantheonPlus}, \rm{Union3}, \rm{DESY5}\}$ as discussed above.
Effects on the values of $P^{\rm{exp}}_{m}(k_{i})$  due to a change of dataset (rather than a change of DDE model), 
expected to be less than a percent level, is not available so far,
which will be neglected for simplicity.  

Besides \cite{DESI:2025xzu}, Ref.\cite{DESI:2024hhd} carried out an earlier DESI Full-Shape analysis on the matter power spectrum.
They differ with each other in at least two points.
First,  Ref.\cite{DESI:2024hhd} adopted a perturbative theory of galaxy power spectrum, 
unlike the effective field theory used by \cite{DESI:2025xzu} within the $k$ range of $0.02-0.2$ h/Mpc.
Second, Ref.\cite{DESI:2024hhd} presented the constraints on cosmological parameters rather than matter power spectrum,
which cannot be directly used to combine the DESI Full-Shape data with other datasets being different from those of \cite{DESI:2024hhd}.

\begin{widetext}
\begin{table*}[]
	\centering
	\resizebox{\textwidth}{!}{
		\begin{tabular}{|c|c|c|c|c|c|c|}
			\hline
			
			&   \multicolumn{2}{c}{Planck+DESI DR2 BAO+PantheonPlus}&   \multicolumn{2}{|c|}{Planck+DESI DR2 BAO+Union3} &   \multicolumn{2}{|c|}{Planck+DESI DR2 BAO+DESY5}\\
			\hline
			
			DESI Y1 & no & yes & no & yes & no & yes \\
			\hline
		 $\omega_\mathrm{c}$
   & $0.11883(0.11927)\pm 0.00093$
   & $0.11926(0.11948)^{+0.00068}_{-0.00049}$
   & $0.11906(0.11939)\pm 0.00087$
   & $0.11922(0.11925)^{+0.00061}_{-0.00052}$
   & $0.11902(0.11934)\pm 0.00084$
   & $0.11965(0.11972)\pm 0.00058$
   \\
   $\omega_\mathrm{b}$
   & $0.02224(0.02226)\pm 0.00014$
   & $0.02220(0.02219)\pm 0.00013$
   & $0.02223(0.02222)\pm 0.00013$
   & $0.02221(0.02222)\pm 0.00012$
   & $0.02223(0.02217)\pm 0.00013$
   & $0.02216(0.02216)\pm 0.00012$
   \\
   $\ln(10^{10}A_s)$
   & $3.0504(3.0471)^{+0.0096}_{-0.012}$
   & $3.0483(3.0440)^{+0.0074}_{-0.012}$
   & $3.0479(3.0392)^{+0.0099}_{-0.012}$
   & $3.0426(3.0420)^{+0.0079}_{-0.011}$
   & $3.049(3.049)^{+0.010}_{-0.011}$
   & $3.0474(3.0453)\pm0.0095$
   \\
   $n_\mathrm{s}$
   & $0.9658(0.9668)\pm 0.0039$
   & $0.9667(0.9661)\pm 0.0041$
   & $0.9654(0.9647)\pm 0.0038$
   & $0.9665(0.9650)\pm 0.0033$
   & $0.9653(0.9642)\pm 0.0037$
   & $0.9652(0.9654)\pm 0.0033$
   \\
   $H_\mathrm{0}$
   & $67.52(67.45)\pm 0.61$
   & $67.40(67.64)\pm 0.62$
   & $66.19(66.27)^{+0.82}_{-1.1}$
   & $65.76(65.73)\pm 0.84$
   & $66.79(66.58)\pm 0.57$
   & $66.66(66.84)\pm 0.56$
   \\
   $\tau_\mathrm{reio}$
   & $0.0593(0.0570)^{+0.0051}_{-0.0066}$
   & $0.0583(0.0573)^{+0.0040}_{-0.0067}$
   & $0.0581(0.0545)^{+0.0052}_{-0.0063}$
   & $0.0557(0.0565)^{+0.0042}_{-0.0058}$
   & $0.0584(0.0590)^{+0.0052}_{-0.0062}$
   & $0.0578(0.0562)^{+0.0048}_{-0.0055}$
   \\
   $w_{0}$
   & $-0.845(-0.840)\pm 0.060$
   & $-0.837(-0.837)^{+0.057}_{-0.051}$
   & $-0.712(-0.709)^{+0.14}_{-0.081}$
   & $-0.671(-0.681)^{+0.091}_{-0.082}$
   & $-0.769(-0.739)^{+0.070}_{-0.059}$
   & $-0.741(-0.759)\pm 0.056$
   \\
   $w_{a}$
   & $-0.58(-0.61)\pm 0.23$
   & $-0.62(-0.65)^{+0.18}_{-0.21}$
   & $-0.93(-0.97)^{+0.28}_{-0.44}$
   & $-1.04(-0.99)^{+0.25}_{-0.29}$
   & $-0.79(-0.91)\pm 0.25$
   & $-0.92(-0.87)\pm 0.20$
   \\   
   \hline	
		\end{tabular}
	}
	\caption{The 68\% credible intervals and the best-fit values (in parentheses) for the $w_{0}w_{a}$CDM model parameters, with and without the DESI Y1 Full-Shape data respectively. }
	\label{values}
\end{table*}
\end{widetext}

\section{Results}
\label{results}
Instead of $\Lambda$CDM with a constant equation of state (EOS) $w=-1$,
we consider the CPL parametrization \cite{Chevallier:2000qy,Linder:2002et} on the EOS:
\begin{eqnarray}{\label{par}}
w(a)=w_{0}+w_{a}(1-a),
\end{eqnarray}
where $\omega_{0}$ and $\omega_{a}$ are two constants.

Table.\ref{values} summarizes the 68\% credible intervals and the best-fit values  (in parentheses) for the $w_{0}w_{a}$CDM model parameters 
in Eq.(\ref{par}) when the Planck+DESI DR2 BAO+SNe data is combined with the DESI Y1 Full-Shape data.
One finds that the new constraints on $w_{0}$ and $w_{a}$ improve over those without the DESI Y1 data. 
In the following we explicitly discuss these results case by case.

%%%%%%%%%%%%%%%%%%%%%%%%%%%%%%%%%%%%%%%%%%%%%%%%
\begin{figure}[htb!]
\centering
\includegraphics[width=8cm,height=8cm]{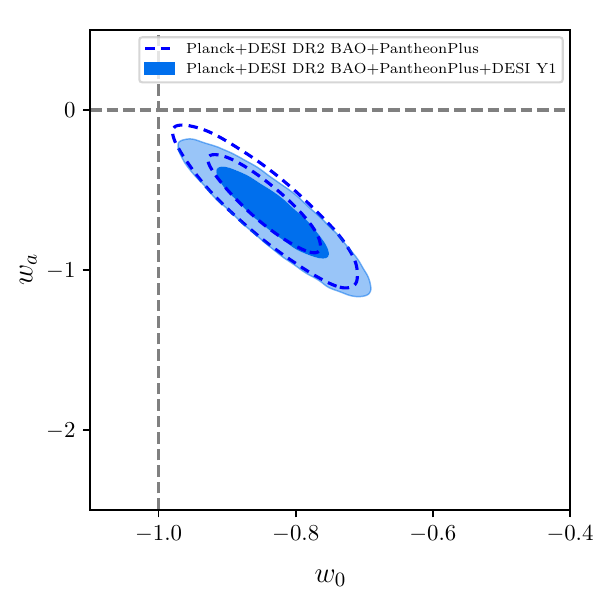}
\centering
\caption{Constraints on $w_{0}$ and $w_{a}$ in Eq.(\ref{par}).
The contours represent the 68\% and 95\% credible intervals.
 The solid (dashed) contours represent the dataset of Planck$+$ DESI DR2 BAO$+$PantheonPlus$+$DESI Y1 with (without) the DESI Y1 Full-Shape data, showing that an inclusion of the DESI Full-Shape data slightly improves the constraints on the two parameters.}
\label{set1}
\end{figure}
%%%%%%%%%%%%%%%%%%%%%%%%%%%%%%%%%%%%%%%%%%%%%%%%

\subsection{Dataset 1: Planck$+$ DESI DR2 BAO $+$ DESI Y1 $+$PantheonPlus}
With the DESY Y1 data added to Planck$+$ DESI DR2 BAO $+$PantheonPlus, 
our MCMC analysis gives 
\begin{eqnarray}{\label{fit1}}
w_{0}=-0.837^{+0.057}_{-0.051},~~w_{a}=-0.62^{+0.18}_{-0.21},~~ \rm{at}~68\%~\rm{CL},\nonumber\\
\end{eqnarray}
as shown in Table.\ref{values}.
Compared to $w_{0}=-0.845\pm 0.060$ and $w_{a}=-0.58\pm 0.23$ at 68\% CL without the DESY Y1 data,
the constraints  in Eq.(\ref{fit1}) get slightly tighter,
with the trend of further running away from the $\Lambda$CDM as seen in Fig.\ref{set1}.

\subsection{Dataset 2: Planck$+$ DESI DR2 BAO $+$DESI Y1$+$Union3} 
\begin{figure}
\centering
\includegraphics[width=8cm,height=8cm]{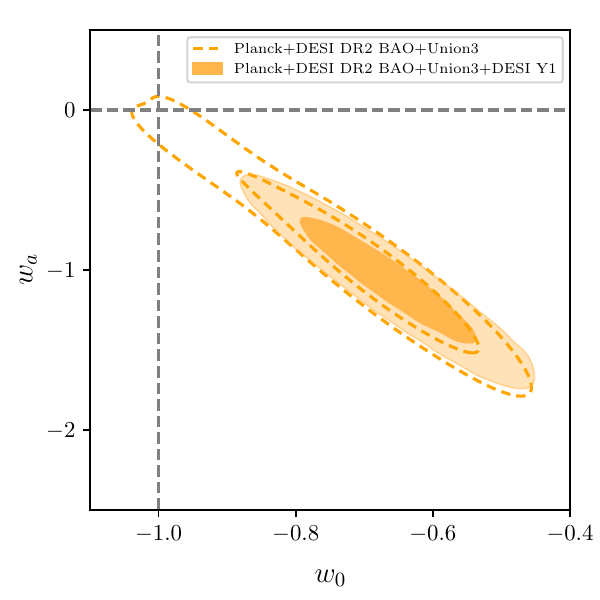}
\centering
\caption{Same as in Fig.\ref{set1} but with the dataset of Planck$+$DESI DR2 BAO$+$Union3,
where the constraints on $w_{0}$ and $w_{a}$ are improved more obviously than in the previous dataset.}
\label{set2}
\end{figure}

Similar to the previous dataset, we combine the dataset of Planck$+$ DESI DR2 BAO $+$Union3 with the DESI Y1 data,
which gives 
\begin{eqnarray}{\label{fit2}}
w_{0}=-0.671^{+0.091}_{-0.082},~~w_{a}=-1.04^{+0.25}_{-0.29},~~ \rm{at}~68\%~\rm{CL},\nonumber\\
\end{eqnarray}
as shown in Table.\ref{values}.
These constraints are more obviously improved over the results  without the DESI Y1 data: 
$w_{0}=-0.712^{+0.14}_{-0.081}$ and  $w_{a}=-0.93^{+0.28}_{-0.44}$ at 68\% CL;
see Fig.\ref{set2} for a comparison.

%%%%%%%%%%%%%%%%%%%%%%%%%%%%%%%%%%%%%%%%%%%%%%%%
%%%%%%%%%%%%%%%%%%%%%%%%%%%%%%%%%%%%%%%%%%%%%%%%

\subsection{Dataset 3: Planck$+$ DESI DR2 BAO $+$ DESY5$+$DESI Y1}

\begin{figure}
\centering
\includegraphics[width=8cm,height=8cm]{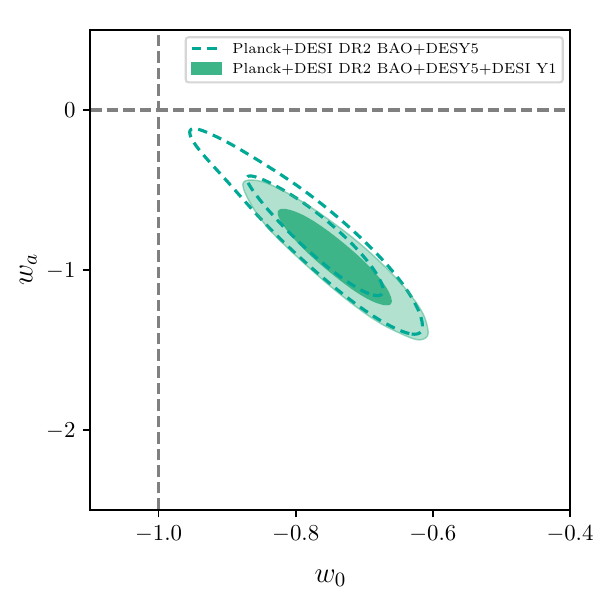}
\centering
\caption{Same as in Fig.\ref{set1} but with the dataset of Planck$+$DESI DR2 BAO$+$DESY5,
where the improvement on $w_{0}$ and $w_{a}$ is the most significant among the three datasets considered.}
\label{set3}
\end{figure}

Finally, we present the MCMC analysis on the dataset of Planck$+$ DESI DR2 BAO $+$ DESY5$+$DESI Y1,
resulting in 
\begin{eqnarray}{\label{fit3}}
w_{0}=-0.741\pm 0.056,~~w_{a}=-0.92\pm 0.20,~~ \rm{at}~68\%~\rm{CL},\nonumber\\
\end{eqnarray}
as shown in Table.\ref{values}.
Relative to the results without the DESI Y1 data: $w_{0}=-0.769^{+0.070}_{-0.059}$ and  $w_{a}=-0.79\pm 0.25$ at 68\% CL,
the improvement is the most significant among the three datasets considered, as shown by Fig.\ref{set3}.

%%%%%%%%%%%%%%%%%%%%%%%%%%%%%%%%%%%%%%%%%%%%%%%%
%%%%%%%%%%%%%%%%%%%%%%%%%%%%%%%%%%%%%%%%%%%%%%%%

\section{Conclusions}
\label{con}
In this study we have combined the DESI DR2 BAO and Full-Shape data to 
place tighter constraints on the $w_{0}w_{a}$CDM, preferred by the DESI BAO data.
Using the DESI Y1 data about the matter power spectrum, 
extracted from the DESI Full-Shape data,
we have built the new DESI Y1 likelihood.
Plugging this DESI Y1 likelihood into three different datasets widely considered in the literature,
we have shown that the constraints on $w_0$ and $w_{a}$ with DESI Y1 data included are improved over  
those without it, especially in the case of DESY5 sample. 

Our method can be applied to at least two aspects. 
First, the constraints on $w_0$ and $w_{a}$ can be further updated by the new DESI Full-Shape data in the next coming years. Second, it is straightforward to repeat our analysis on the DDE model under parameterizations rather than 
the CPL adopted.

\begin{acknowledgements} 
Q. Zhou thanks R. Cereskaite for correspondence about the DESI Y1 data.
\end{acknowledgements}

\end{document}